\documentclass[journal]{IEEEtran}
\usepackage{comment}
\usepackage{hhline}
\usepackage{scrextend}
\usepackage{color}
\usepackage{mathtools}
\usepackage{verbatim}
\usepackage{subcaption}
\usepackage{float}
\usepackage{multirow}

\begin{document}

\title{Hotspot-aware DSA Grouping and Mask Assignment}
%
%
%
%
%
\author{Yasmine~Badr and 
        Puneet~Gupta 
        \\ybadr@ucla.edu, puneet@ee.ucla.edu
        \\Electrical and Computer Engineering Department
        \\University of California, Los Angeles}
\maketitle
\begin{abstract}
In Directed Self Assembly (DSA), poor printing of guiding templates can cause misassembly resulting in high defect probability. Therefore, hotspots should be avoided in the choice of the DSA groups. 
Accordingly, Directed Self-Assembly (DSA) technologies which use Multiple Patterning (MP) to print the guiding templates need to be aware of hotspots during the DSA grouping and MP Decomposition. In this paper, we present a hotspot-aware heuristic for DSA grouping and MP decomposition. Results show that that the proposed heuristic eliminates 78\% of the hotspots and conflicts that result from using a hotspot-unaware grouping and decomposition algorithm. In comparison to the optimal solution using Integer Linear Programming, the proposed heuristic results in ~24\% more violations.
\end{abstract}


\section{Introduction}
\begin{figure}[h]
\centering
\includegraphics [width=0.5\textwidth]{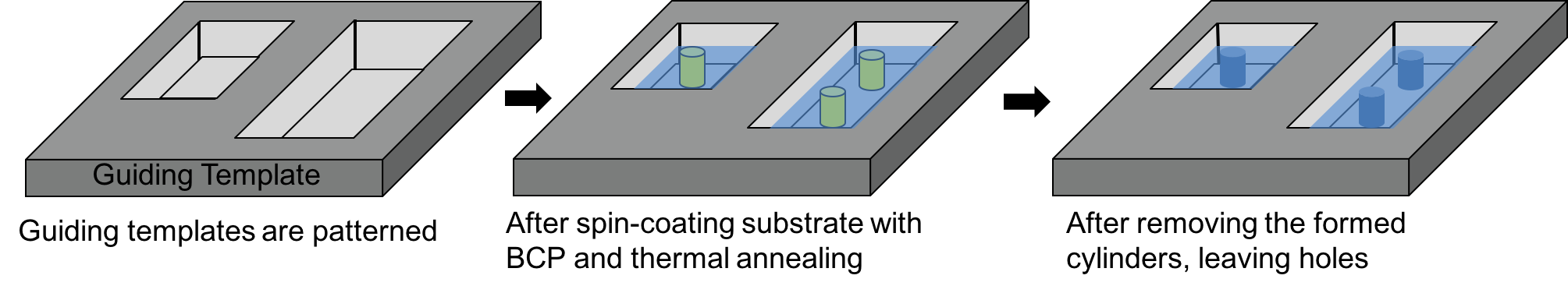}
\caption{DSA process for contact/via holes.
\label{fig:selfAssemblyHotspot}}
\end{figure}

Directed Self Assembly (DSA) is a promising patterning technique for the sub-7nm nodes because of its inherent pitch multiplication features and low cost ~\cite{dsaBossung}. There are two main types of DSA: graphoepitaxy and chemoepitaxy, but we focus on graphoepitaxy because it is more appropriate for patterning of random features required for via/contact layers ~\cite{gronheid2016euv}, which are of interest in this work.  Graphoepitaxy process is shown in Figure ~\ref{fig:selfAssemblyHotspot}. First, the guiding templates are patterned using a lithography technique, then the Block Copolymer (BCP) undergoes annealing and self-assembles into regular structures (cylindrical formations in this example). This way, via holes with a pitch smaller than that allowed by the lithography technique can be realized, as shown in the case of the two neighboring holes in the same guiding template in Figure ~\ref{fig:selfAssemblyHotspot}.

\subsection{Why Hybrid DSA-MP?}
Earlier work has shown that using DSA with Multiple Patterning (MP) can reduce the cost of the process by reducing the number of required masks ~\cite{ma2014challenges}. In addition, DSA is unlikely to achieve a sub-7nm node without MP, due to the need for higher resolution to print the guiding templates.

\subsection{What is considered a hotspot?}
A hotspot can be one of the following:
\begin{enumerate}
\item {Lithographic hotspot. This is a low-yield pattern due to photolithography, which is likely to cause a printing failure \cite{ding2011high,badr2017technology}. }
\item{Complex design rule. In advanced nodes, foundries introduced a lot of complex 2D and conditional rules. These rules can require pattern-based representation \cite{badr2017technology, DRCPlus}}
\item{Forbidden pattern due to using a restrictive patterning technology like Self-Aligned Multiple Patterning \cite{badr2017technology}.}
\item{Some guiding templates result in high probability of defects in self-assembly. Thus the prohibition of low yield DSA groups or guiding templates can be achieved by representing these guiding templates as hotspots \cite{badr2017technology,shim2015defect}.}
\end{enumerate}
\subsection{Prior Work in DSA-MP}
Several works \cite{dsaMPSPIE14,DSA_MP_DAC15,Kuang2016,Xiao2016} have addressed the problem of simultaneous DSA Grouping and Mask assignment for hybrid DSA-MP technologies.  Ou et al. \cite{Ou2016} solved the same problem while adding redundant vias, while Lin et al. \cite{Lin2016} added cut redistribution. DSA-aware routing has been addressed in \cite{ou2017dsar} for DSA+Double Patterning (DP) technology. In ~\cite{shim2015defect}, the authors perform DSA-friendly post-placement optimization, to avoid DSA groups of high defect probability, for contacts layer. However, they do not address low yield patterns made up of a neighborhood of DSA groups. 
The problem of Hotspot-aware DSA Grouping and Mask Assignment was considered in  ~\cite{badr2017technology} for the purpose of Technology Path-finding, and an Integer Linear Program (ILP)  was used for optimal evaluation, which is not scalable for full-chip layouts.
\subsection{Contribution}
\begin{figure}[ht]
\centering
\includegraphics [width=0.5\textwidth]{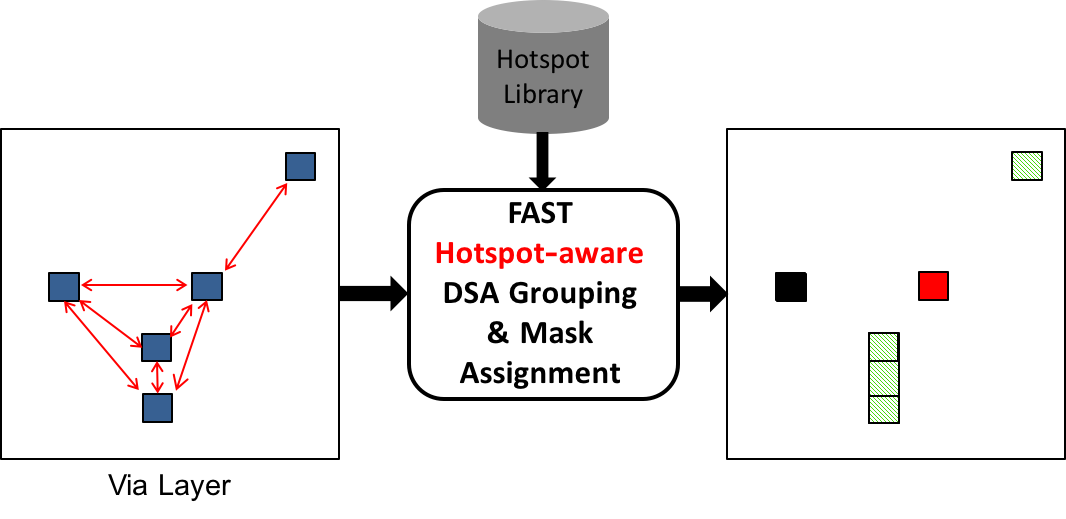}
\caption{Objective of this work. In this example, DSA+Triple Patterning is assumed.
\label{fig:hsObjective}}
\end{figure}

In this work, we propose a scalable heuristic to achieve hotspot-aware DSA grouping and MP decomposition for hybrid DSA + MP technologies where MP is used to pattern the DSA guiding templates. Given, a contacts/via layer, it is required to cluster the vias into DSA groups, where each group will result in a guiding template, and assign the groups to the different masks, while avoiding \textbf{hotspots} as shown in Figure \ref{fig:hsObjective}.  

We propose a heuristic which can be applied with different DSA Grouping and MP decomposition algorithms. Therefore, one key advantage of this heuristic is that it doesn't disrupt the flow of DSA Grouping and MP decomposition, but rather can be integrated nicely into the several threads of research coming up with fast methods for DSA grouping and MP decomposition. Up to the authors' knowledge, this is the first work which presents a scalable method for considering hotspots in the DSA Grouping and MP decomposition problem.

The rest of the paper is organized as follows. Section \ref{sec:hsAvoid} describes how the hotspot avoidance algorithm is mapped to a Set Cover problem and solved. Section \ref{sec:hsResult} explains the experiments and results, followed by the conclusion in Section \ref{sec:hsConcl}.

\section{The Hotspot-avoidance Heuristic\label{sec:hsAvoid}}
We choose to design the hotspot avoidance algorithm as a pre-processing algorithm before performing the mask assignment and DSA grouping. The hotspot avoidance problem is mapped to the classic Set Cover problem, and we use the greedy heuristic to solve it. In this section, we explain the necessary details of mapping the problem from hotspot avoidance to the Set Cover problem.

\subsection{Pattern and Graph Representation}
\begin{figure}[ht]
\centering
\includegraphics[width=0.5\textwidth]{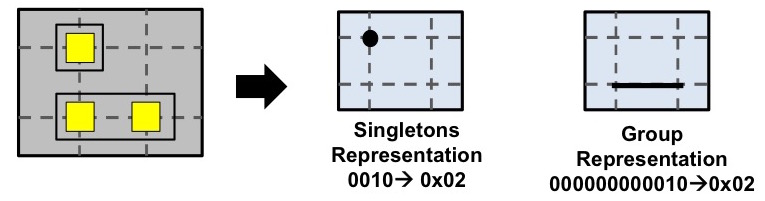}
\caption{A hotspot and its corresponding representation ~\cite{badr2017technology}
\label{fig:hotSpotRep}}
\end{figure}

We use the pattern representation proposed in ~\cite{badr2017technology}, shown here for convenience in Figure ~\ref{fig:hotSpotRep}. The segment representation encodes the DSA groups in the pattern while the node representation expresses the singletons (DSA groups containing one via each) in the pattern. 

For a  hotspot pattern and a layout window, ``\textbf{constituent vias}'' are the vias, in the layout window, which appear in DSA groups or singletons in the corresponding locations in the hotspot pattern. Vias which lie in the layout window but are not constituent vias are simply called ``\textbf{non-constituent vias}''. For example, for the layout window and the hotspot pattern shown in Figure ~\ref{fig:graphModHotspotAvoid},  vias $a$, $b$ and $d$ are ``\textbf{constituent vias}'' while via $c$ is a ``\textbf{non-constituent via}''. 

In the following, ``\textbf{potential hotspot}'' is a layout window where a hotspot from the hotspots library can possibly exist, after grouping and coloring of the vias in the window. This means that the prerequisites of a potential hotspot is that for every via in the hotspot pattern, there is a via in the layout window in the corresponding location. For example in Figure ~\ref{fig:graphModHotspotAvoid}, the shown layout window is a potential hotspot; because if vias \textit{a} and \textit{b} are grouped and assigned to the same mask as  \textit{d} and if via \textit{c} is assigned to a different mask, the hotspot shown in the same figure will occur in this layout window. More about that will be described in Section \ref{sec:elimPotHS}.

We assume the layout is initially represented using a hybrid hyper-graph, where a conflict hyper-edge/edge is created between every two vias separated by a distance less than the minimum allowed distance on a single mask, and a grouping hyper-edge is created between every set of vias which can form a legal DSA group.

\subsection{How to eliminate potential hotspots?\label{sec:elimPotHS}}
\begin{figure}[ht]
\centering
\includegraphics[width=0.5\textwidth]{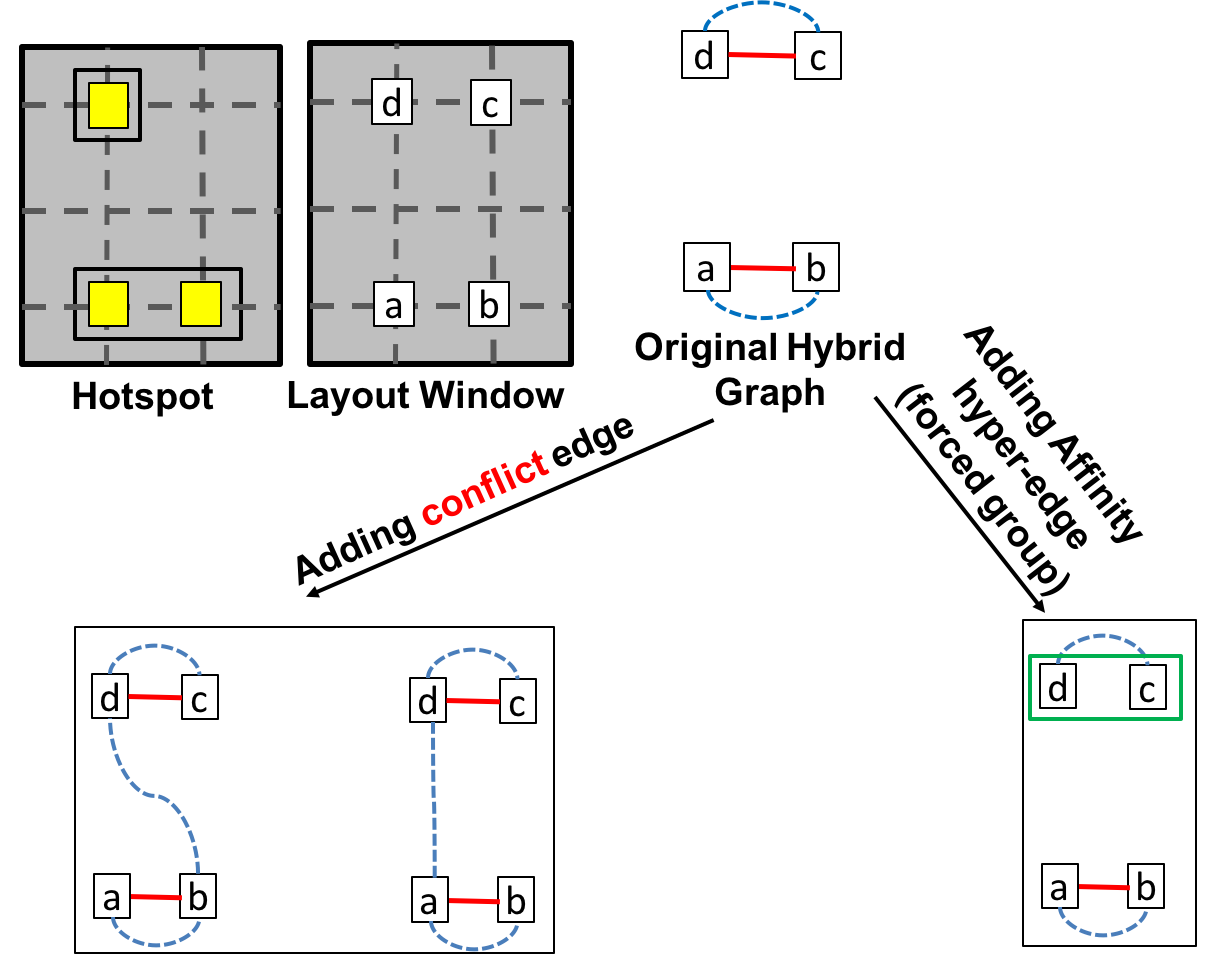}
\caption{Elimination of the potential hotspots
\label{fig:graphModHotspotAvoid}}
\end{figure}

 A potential hotspot will result in the presence of a hotspot in the output masks if:
\begin{enumerate}
\item{all the constituent vias are assigned to the same mask, and}
\item{none of the non-constituent vias is assigned to that mask.} 
\end{enumerate}
 For example, the hotspot in Figure ~\ref{fig:graphModHotspotAvoid} will not exist in the shown layout window unless vias \textit{a} and \textit{b} are grouped and assigned to the same mask as \textit{d}, and  via \textit{c} is assigned to a different mask. Thus, to avoid this hotspot in the shown layout window, a conflict edge can be added between vias \textit{b} and \textit{d} to force them onto different masks, or between vias \textit{a} and \textit{d}. Alternatively, the hotspot can be avoided if vias \textit{c} and \textit{d} are grouped (which of course means that they are assigned to the same mask).\footnote{We assume that if two vias are within a grouping distance from each other and are assigned to the same mask, they are definitely grouped.}

 Thus, eliminating a potential hotspot can be done by one of the following two ways:
 \begin{enumerate}
 \item{Adding a \textbf{conflict edge} between two of the constituent vias of the potential hotspot. The two chosen constituent vias must not be group-able. }
 \item{Forcing a DSA group between at least one constituent via and one non-constituent via. We use ``\textbf{affinity hyper-edge}'' to refer to a forced DSA group.}
 \end{enumerate}
 
 However, adding a conflict edge or forcing a DSA group adds additional constraints to the grouping and coloring problem, which can increase the number of unresolved conflicts. Thus, it is desired to eliminate the hotspots by adding the \textbf{minimum} number of conflict edges and forced groups.
\subsection{Mapping the problem to a Set Cover problem}
A hotspot can be eliminated by adding one edge from a set of conflict edges, or forcing a group represented as an affinity hyper-edge between vias. In some cases, one conflict or affinity hyper-edge can be used to eliminate more than one potential hotspot. The objective is to choose the minimum number of edges which cover the potential hotspots.  This is the classic Set Covering problem ~\cite{karp1972reducibility} where each set contains all the potential hotspots that can be eliminated by the addition of  a particular conflict  or an affinity hyper-edge. This means that there is one set corresponding to each potential conflict edge and each affinity hyper-edge. The universe of elements is made of all the potential hotspots.

\subsection{Greedy Heuristic to solve the  Set Covering problem}
\begin{figure}[ht]
\centering
\includegraphics[width=0.25\textwidth]{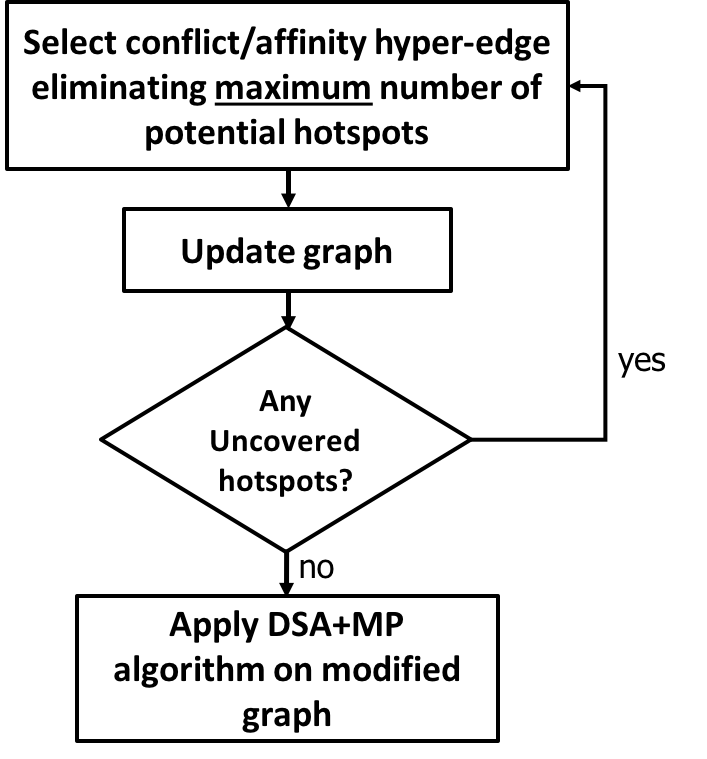}
\caption{Hotspot avoidance Flow using Greedy Set Cover heuristic
\label{fig:hsFlow}}
\end{figure}

Set Covering is an NP-complete problem. We use the Greedy heuristic to solve the Set Covering problem \cite{Johnson1974}. The algorithm is to iteratively choose the subset covering the largest number of elements from the universe set. Applying the greedy heuristic to the Hotspot avoidance problem results in the flow shown in Figure ~\ref{fig:hsFlow}, where in each iteration the conflict  or affinity hyper-edge which eliminates the largest number of potential hotspots is chosen. The graph is updated according to the chosen hotspot avoidance method. Finally the DSA grouping and mask assignment is performed on the updated group.
\subsection{Implementation}
We develop an $O(m)$ implementation by using the bucket list data structure similar to the one used in ~\cite{fiduccia1988linear}, where m is the number of potential hotspots.
For every potential hotspot, we identify the pairs of constituent vias that are not groupable and do not have a conflict edge between them. These are potential to-add conflict edges. Furthermore, the potential DSA groups including a constituent via and a non-constituent via are identified. These are potential forced groups.

Each edge or forced group is attached to the list of edges of the same frequency, i.e. the number of potential hotspots that will be avoided by this edge. Frequencies are saved in a linked list in decreasing order. Then in each iteration in Figure ~\ref{fig:hsFlow}, one edge from the highest frequency is picked in $O(1)$.
\subsection{Which DSA Grouping and MP Decomposition algorithm can be used? }

As shown in Figure \ref{fig:graphModHotspotAvoid}, the proposed hotspot heuristic adds conflict edges as well as forced groups. Handling an added conflict edge is straightforward, since it is to be exactly considered as an original conflict edge ( that exists because the distance between two vias is smaller than the minimum allowed distance on a mask).  

Some algorithms can handle forced groups easily. In \cite{kuang2016simultaneous}, the forced groups are to be modeled by assigning a negative cost to the grouping edges representing them. Finally, the algorithm in ~\cite{ou2017efficient} can handle the forced groups very simply by setting the ILP  variables corresponding to the selected groups. 


\section {Experiments and Results\label{sec:hsResult}}

\subsection{Generating Hotspots}
\begin{figure}[ht]
\centering
\includegraphics[height=0.1\textheight]{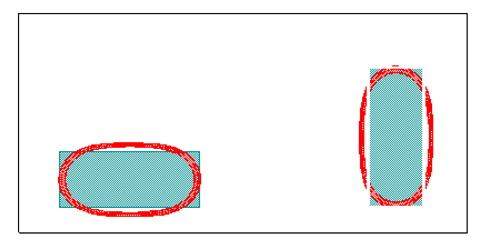}
\caption{PV Band in red
\label{fig:pvband}}
\end{figure}

In order to generate hotspots, we randomly generated 4x2 patterns (four rows and two columns). 
The randomly generated patterns  were then simulated using Calibre Litho-Friendly Design tool \cite{LFD} to generate the Process Variation (PV) bands shown in Figure ~\ref{fig:pvband}. We use the Process Manufacturability Index (PMI) metric proposed in \cite{torres2005integrated} to gauge the sensitivity of each pattern to process variation. Patterns with PMI higher than 20\% were considered as hotspots and there were 36 hotspots.
\subsection{Results}
\subsubsection*{Optimal ILP formulation for Hotspot-aware Grouping and Coloring}
In ~\cite{badr2017technology}, an ILP formulation was proposed for hotspot-aware simultaneous DSA Grouping and MP decomposition. We apply the same formulation in order to find the optimal solution and benchmark the developed heuristics against the optimal solution. 

\subsubsection*{Experimental Setup}
We synthesized, placed and routed the following designs: AES \footnote{In order to compare to the optimal solution using ILP, we had to use a clip of AES instead of the full layout because otherwise the ILP needed more than 24 hours } and MIPS from OpenCores, and ARM Cortex M0 using a projected 7nm library that was scaled down to 5nm. The hotspot heuristic is implemented in C++. OpenAccess is used for layout manipulation and Boost Graph API is used for the graph operations. Each connected component of the graph representation of the layout is processed independently for the hotspot heuristic and for grouping and coloring. We used OpenMP threads for \textbf{parallel} processing of the connected components. The experiments were run on a computing cluster with 48G of virtual memory, using four threads.

We apply the greedy set cover heuristic followed by DSA grouping and Mask Assignment. For grouping and coloring, we used the ILP in \cite{badr2017technology} -without providing a hotspot library- in order to purely measure the quality of the hotspot-related heuristic independently from the quality of the grouping and coloring heuristics. %
To use the ILP of \cite{badr2017technology} for grouping and coloring only, we do not provide any hotspot library. 
We only used collinear DSA groups, assuming 193i is used to pattern the guiding templates \footnote{The same hotspot heuristic can be used if the set of allowed DSA groups is different, and hence can be used if Extreme Ultraviolet Lithography (EUV) is used instead of 193i.}, and used the parameter values shown in Table ~\ref{tab:hsParam}.

\begin{table}[ht]
\centering
\caption{{Parameter values used for ILP in ~\cite{badr2017technology} for DSA grouping and Mask Assignment}\label{tab:hsParam}}
{
\begin{tabular}{|l||l|}
\hline 
\textbf{Parameter} & \textbf{Value}\tabularnewline
\hline 
$L_0$ & 30 nm\tabularnewline
\hline 
$max\_dsa$\_pitch & 51 nm\tabularnewline
\hline 
$max\_g$ & 2 or 3\tabularnewline
\hline
$min\_pitch\_same\_mask$ & 75 nm\tabularnewline
\hline
$min\_pitch\_diff\_mask$ & 10 nm\tabularnewline
\hline
$via\_width$ & 15nm\tabularnewline
\hline 
\end{tabular}}
\end{table}

We compare our results to the ILP formulation for hotspot-aware DSA grouping and coloring in ~\cite{badr2017technology}. We also compare the results of the hotspot heuristic to the results of using an optimal DSA grouping and coloring algorithm which is unaware of hotspots, and for that we use the ILP of ~\cite{badr2017technology} \textbf{without} using any hotspots library as input to the ILP. The results are shown in Tables ~\ref{tab:resultsHS_g2} and ~\ref{tab:resultsHS_g3} for maximum group size=2 and maximum group size=3 respectively. The shown number of violations is the sum of number of unresolved conflicts as well as number of existing hotspots. 

The Greedy set cover heuristic produces a total of 24\% increase in violations in comparison to the optimal solution. In comparison to the optimal grouping and coloring algorithm that is unaware of hotspots, our hotspot heuristic eliminated 78\%  of the total number of conflicts and hotspots in the output masks.

The average runtime of the heuristic (excluding the DSA grouping and MP assignment time)  is 45 seconds for the used test cases.

\begin{table}[h]
\centering
\caption{Results of Greedy Set cover heuristic using maximum group size=2\label{tab:resultsHS_g2}}
\begin{tabular}{|c|c|c|c|c|c|}
\hline 
 & \multirow{2}{*}{N Vias} & DSA\_Pathfind \cite{badr2017technology} & hotspot-unaware & \multicolumn{2}{|c|}{Our heuristic}\tabularnewline
\cline{3-6}
& & N Viols & N Viols & N Viols &Time(s)\tabularnewline
\hline 
aes\_clip & 35051 & 71 & 350&79 & 36\tabularnewline
\hline 
mips & 35971 & 12 &129& 21 & 73\tabularnewline
\hline 
m0 & 44530 & 65 & 280&74 & 34\tabularnewline
\hline 
Ratio & -- & 1 & 5.13&1.18 & --\tabularnewline
\hline 
\end{tabular}
\end{table}

\begin{table}[h]
\centering
\caption{Results of Greedy Set cover heuristic using maximum group size=3\label{tab:resultsHS_g3}}
\begin{tabular}{|c|c|c|c|c|c|}
\hline 
 & \multirow{2}{*}{N Vias} & DSA\_Pathfind \cite{badr2017technology} & hotspot-unaware & \multicolumn{2}{|c|}{Our heuristic}\tabularnewline
\cline{3-6}
& & N Viols & N Viols & N Viols &Time(s)\tabularnewline
\hline 
aes\_clip & 35051 & 50 & 292 & 63 & 36\tabularnewline
\hline 
mips & 35971 & 12 & 130& 20& 35\tabularnewline
\hline 
m0 & 44530 & 39 &237& 52 & 57\tabularnewline
\hline 
Ratio & -- & 1 & 6.52& 1.34 & --\tabularnewline
\hline 
\end{tabular}

\end{table}

\section{Conclusion\label{sec:hsConcl}}
We proposed an algorithm for Hotspot-aware DSA Grouping and MP decomposition. The proposed heuristic was benchmarked against the optimal solution, resulting in ~24\% more violations. In comparison to an optimal DSA grouping and mask assignment algorithm that is unaware of hotspots, the proposed heuristic eliminates ~78\% of the conflicts and hotspots.
\section{Acknowledgments}
This work has been partially funded by NSF and CDEN (http://cden.ucsd.edu).

%
\bibliographystyle{IEEEtran}
\bibliography{references}  
%

\end{document}